4# Rapid Transport of Glassy Supersolid Helium in Wavy-Rough Nanpores

Zotin K.-H. Chu

*3/F, 4, Alley 2, Road Xiushan, Le-Shan-Xin-Cun, Xu-Jia-Hui 200030, PR China***One prominent difference between the transport in cylindrical nanopores（i.e., nanotubes）and those in macroscale pores is the (relatively) larger roughness observed in the wall of nanotubes. As the tube size decreases, the surface-to-volume (S/V) ratio increases. Therefore, surface roughness will greatly affect the transport in cylindrical nanopores. As commented by Rittner and Reppy[1] : An unresolved puzzle is presented by results of Kim and Chan in (porous) Vycor (glass [2]) and porous gold [3]; in both cases, the S/V ratios are orders of magnitude larger than for our (Rittner and Reppy [1]) cells, yet the supersolid fractions ($\rho_s/\rho$) are on the order of 2% (while Rittner and Reppy's $\rho_s/\rho$ is around 20%). Here we show that the presumed wavy roughness distributed along the wall of different nanopores (radius : *a* around 3.5 nm for Vycor [2] or a silica glass; around 245 nm for porous gold [3]) will induce larger volume flow rates of solid helium (of which there is a minimum) which might explain reported experimental differences of the supersolid fractions observed so far.**

As proposed by Andreev [4] and Rittner and Reppy [1] the observed supersolidity might be in a glassy solid (helium) state. Both idea could be inspired from the annealing effect of supersolidity reported before [5]. To consider the transport of this kind of glass or quantum glass [4] in nanodomain, we adopt the verified model initiated by Cagle and Eyring [6] which was used to study the annealing of glass. To obtain the law of annealing of glass for explaining the too rapid annealing at the earliest time, because the relaxation at the beginning was steeper than could be explained by the bimolecular law,

Cagle and Eyring tried a hyperbolic sine law between the shear (strain) rate : $\gamma$ and (large) shear stress : $\tau$ and obtained the close agreement with experimental data. This model has sound physical foundation from the thermal activation process (Eyring [7] already considered a kind of *tunneling* which relates to the matter rearranging by surmounting a potential energy barrier) and thus it could resolve the concern raised by Anderson [8] for the thermal noises to the superflow of vortex liquid. With this model we can associate the (glassy) fluid with the momentum transfer between neighboring atomic clusters on the microscopic scale and reveals the atomic interaction in the relaxation of flow with (viscous) dissipation. Thus the hyperbolic sine law (which reads $\dot{\gamma} = \dot{\gamma}_0 \sinh(\chi r/a)$, with $\chi$=(-a dp/dz)/(2$\tau_0$); $\dot{\gamma}_0, \tau_0$ are function of temperature and have the same dimension as $\dot{\gamma}, \tau$ ; dp/dz is the pressure-gradient along the nanotube-axis direction (z)) can be used to describe rapid transport on the nanoscale.

The main difficulty is, however, to theoretically handle the transport across the wavy roughness distributed along the wall of cylindrical nanopores (or nanotubes) of the porous Vycor glass [2]. If the mean radius of the nanotube is *a* (nm), we presume the wavy roughness to be $r = a + \varepsilon \sin(k\theta)$ where $(r, \theta)$ is the cylindrical coordinate we adopt, $\varepsilon$ is (peak) amplitude of the (wavy) roughness, and *k* is the wave number. The boundary condition is borrowed from Thompson and Troian [9] : $U_{slip} = L_s \dot{\gamma} (1-\dot{\gamma}/\dot{\gamma}_c)^{-1/2}$ (considering the slip-type superflow [10], $L_s$ : slip length, $\dot{\gamma}_c$ : is the critical shear rate at which the slip length diverges). Using the boundary perturbation approach [11], with $\dot{\gamma} = -dU/dr$ and the above boundary condition, we can obtain the volume flow rate (*Q*, up to the second order) after lengthy mathematical manipulations. Note that we should calculate $\dot{\gamma} = -dU/dn$ at the surface $r = a + \varepsilon \sin(k\theta)$ where *n* is the normal to the surface ($dU/dn = \nabla U \cdot \vec{n}$, $\vec{n}$ is an unit vector along the *n*-direction [11]).

The key results are illustrated in Fig. 1. Parameters are selected as : $a$=1 (nm), $\varepsilon$=0.1$a$, $L_s$=$a$, $\dot{\gamma}_c/\dot{\gamma}_0$=10. $\chi/a$ (x-axis) represents the ration of forcing (along the z-direction) per unit volume and (referenced) shear stress. We can observe the larger (flow-rate) differences between the approximately smooth and wavy-rough cylindrical nanopores for small $\chi/a$ ( $\leq 0.1$) situations. In fact, for silicon plates [12] roughness-peak could be around 0.2 nm! To be specific, the maximum velocity we calculated (using above parameters) inside the nanopore is around 400 (nm/s) for $\chi/a \approx 5$ (set $\dot{\gamma}_0 =1$ s$^{-1}$). With our results, it seems to us that the almost the same order of 2% for $\rho_s/\rho$ in different nanopores (3.5 and 245 nm) measured by Kim and Chan [2,3] should follow the smooth case in Fig. 1 compared (relatively) to the largely disordered or rough (presumed to be wavy) case [5] where $\rho_s/\rho$ is around 20%. Meanwhile the ratio of $\varepsilon/a$ could be almost the same for both (porous) Vycor [2] and gold [3] samples to account for the same order of measured supersolid fractions (presumed $L_s \sim a$ and the corresponding scaled-down shear rate [12]). However, a supersolid critical velocity ~ 10 μm/s or 10000 nm/s was reported in both systems [2,3] which is too large (25 times larger) compared to our calculations (up to second order). We also noticed that Rittner and Reppy [5] reported the annealing effect : decreasing $\rho_s/\rho$ to almost zero. This annealing process could make transitions from states firstly along the upper (rough) curve and then to the lower curve (smooth cases) in Fig. 1. Other interesting results are for smooth cases : Increasing forcing will enhance the transport a little; while for rough cases : (starting from initially quiescent environment) Increasing forcing might reduce the transport until reach a minimum flow rate as evidenced in Fig. 1. As $\varepsilon \sim 0.01$, the rough case collapses to the smooth case, but once $\varepsilon \sim 0.05$, there are clear differences between the transport for smooth and rough cases. We thus propose that future measurements should provide the information of roughness distribution along the nanopores for settling down the present supersolidity puzzles [13-15].

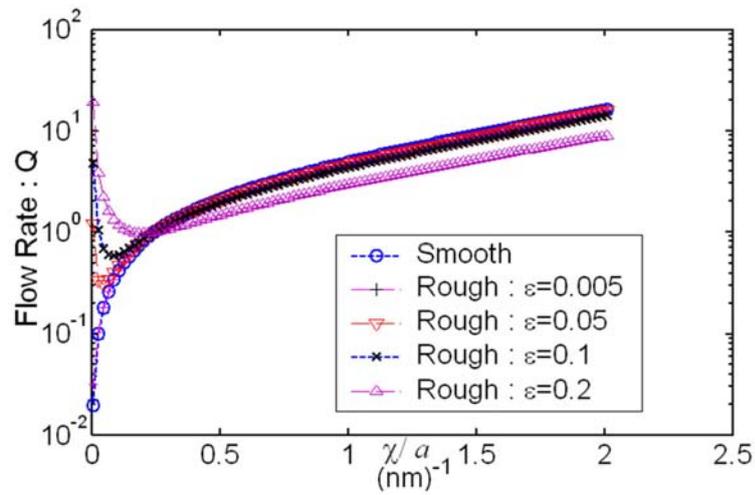

**Figure 1** Enhanced transport of glassy solid helium along wavy-rough (instead smooth) nanopores. $\chi/a$ : forcing (along z-direction) per unit volume or $\chi = (-a\, dp/dz)/(2\tau_0)$. ε is (peak) amplitude of the (wavy) roughness, here the mean radius of cylindrical nanopore : $a$ is set to 1 (nm), the wave number of wavy-roughness $k=10$. Transport along wavy-rough nanpores will have a minimum while that for smooth case will not. There is transport even forcing is absent.